\documentclass[12pt]{amsart} 
\usepackage{amsmath,amsthm,amsfonts,amscd,amsxtra,amsopn,verbatim,hyperref,array}
\usepackage{eucal}
\usepackage{epsfig}
\usepackage{amsfonts}
\usepackage{amsmath}
\usepackage{amsthm}
\usepackage{psfrag}
\usepackage{pdfpages}

\usepackage{mathrsfs}
\usepackage[all]{xy}

\renewcommand{\mathfrak}{\mathcal}
\renewcommand{\mathcal}{\bold}

\usepackage{natbib}
\usepackage{hyperref}

\newcommand{\beq}{\begin{equation}}
\newcommand{\eeq}{\end{equation}}
\newcommand{\beqarr}{\begin{eqnarray}}
\newcommand{\eeqarr}{\end{eqnarray}}
\newcommand{\beqa}{\begin{eqnarray*}}
\newcommand{\eeqa}{\end{eqnarray*}}
\begin{document}

\title[Recursive compartment model]{Proposal of a  recursive compartment model of epidemics and applications to the Covid-19 pandemic}
\author[\small M. Kreck, E. Scholz]{\small  Matthias Kreck$^{\dag}$,  Erhard Scholz $^{\ddag}$}
\date{August 31, 2020}

\begin{abstract} \tiny {This is work in progress. We make it accessible hoping that people might find the idea useful. We propose a discrete, recursive 5-compartment model for the spread of epidemics, which we call {\em  SEPIR-model}.  Under mild assumptions which typically are fulfilled for the Covid-19 pandemic it can be used to reproduce the development of an epidemic from a small number of parameters closely related to the data. We demonstrate this at the development in Germany and Switzerland. It  also allows model predictions assuming nearly constant reproduction numbers. Thus it might be   a useful tool for shedding light on which interventions might be most effective in the future. In future work we will discuss other aspects of the model and  more countries. }
\end{abstract}

\maketitle
\renewcommand{\thefootnote}{\fnsymbol{footnote}}

\footnotetext[2]{Mathematisches Institut der Universit\"at Bonn and Mathematisches Institut der Universit\"at Frankfurt, Germany, \quad kreck@math.uni-bonn.de}
\footnotetext[3]{University of Wuppertal, Faculty of  Math./Natural Sciences, and Interdisciplinary Centre for History and Philosophy of Science, \quad  scholz@math.uni-wuppertal.de}
\renewcommand{\thefootnote}{\arabic{footnote}}

\section{Introduction} We propose a new model for the development of an epidemic, which we call SEPIR model. The name stands for 5 compartments which people pass through in the course of an epidemic, the compartment S of {\em Susceptibles}, E of {\em Exposed}, P of {\em Propagators}, who infect other people, I of  {\em Isolated}, either in quarantine or in hospital,  and R of {\em Removed}. The most important compartment is $I$, this occurs in the data sets under the name ``active cases'', the people which are reported to be infected and are either sent to quarantine or hospital, so we call them isolated. Thus the "I" also stands for those counted as infected (but, since in quarantine or hospital, not infecting others). It is a simple recursive model. We have tried to formulate it such  that also non-mathematicians can read it. The details will be explained in the next section. 

Our motivation is to look for a model, which leads to a good approximation of the data curves. In order not to overload this text with a bulk of graphics and  data we decide to concentrate on the comparison for two countries, Germany and Switzerland. The reason for choosing just these is  that the documentation of the data  by their health organizations  seems to be  comparatively reliable. In future work, which we write together with Harald Grohganz (whom we thank for programming our model, most of the graphics are made with his program), we will discuss many other countries and  elaborate deeper on some aspects of the model and its comparison with real data \cite{GKS}.

For the comparison of a model with real data it is crucial to develop methods which allow to derive the input parameters of the model from the data. We discuss this in section 3. 
A model can be useful only if it is possible to 
distil a small number of parameters  from the data, which allow it to   reconstruct or to predict what happens. There is a  fundamental parameter for modelling an epidemic, the reproduction number. The idea of this number is very simple, namely it is the number  of people that one infected person will pass the virus on to, on average.  It is not easy to read it off from the data (and to our best knowledge there is no unique answer as one can see from different data sources).  In our model we give a way to determine it. If this number were a constant (and other parameters,  which influence the data, too, like for example the duration people are sent to quarantine), then a good model should describe or predict the epidemic by equations making essential  use of this input parameter. But as we all know this number is changing. On the other hand,  one observes that there are more or less longer periods where the reproduction number is approximately constant. During such a period the SEPIR model gives a  good approximation of the development of an epidemic. 

In most countries there are longer periods of this type, so that one can use a small number of corresponding parameters to solve the model equations. In Germany we observe  only 5 such periods between March 25 and August 9,  resulting in 5 different values for the input parameter related to the reproduction number. The following graphics shows the result (a more detailed  description is to be found in the caption of fig. \ref{SEPIR D}):

\hspace*{1cm}
\includegraphics[scale=0.4]{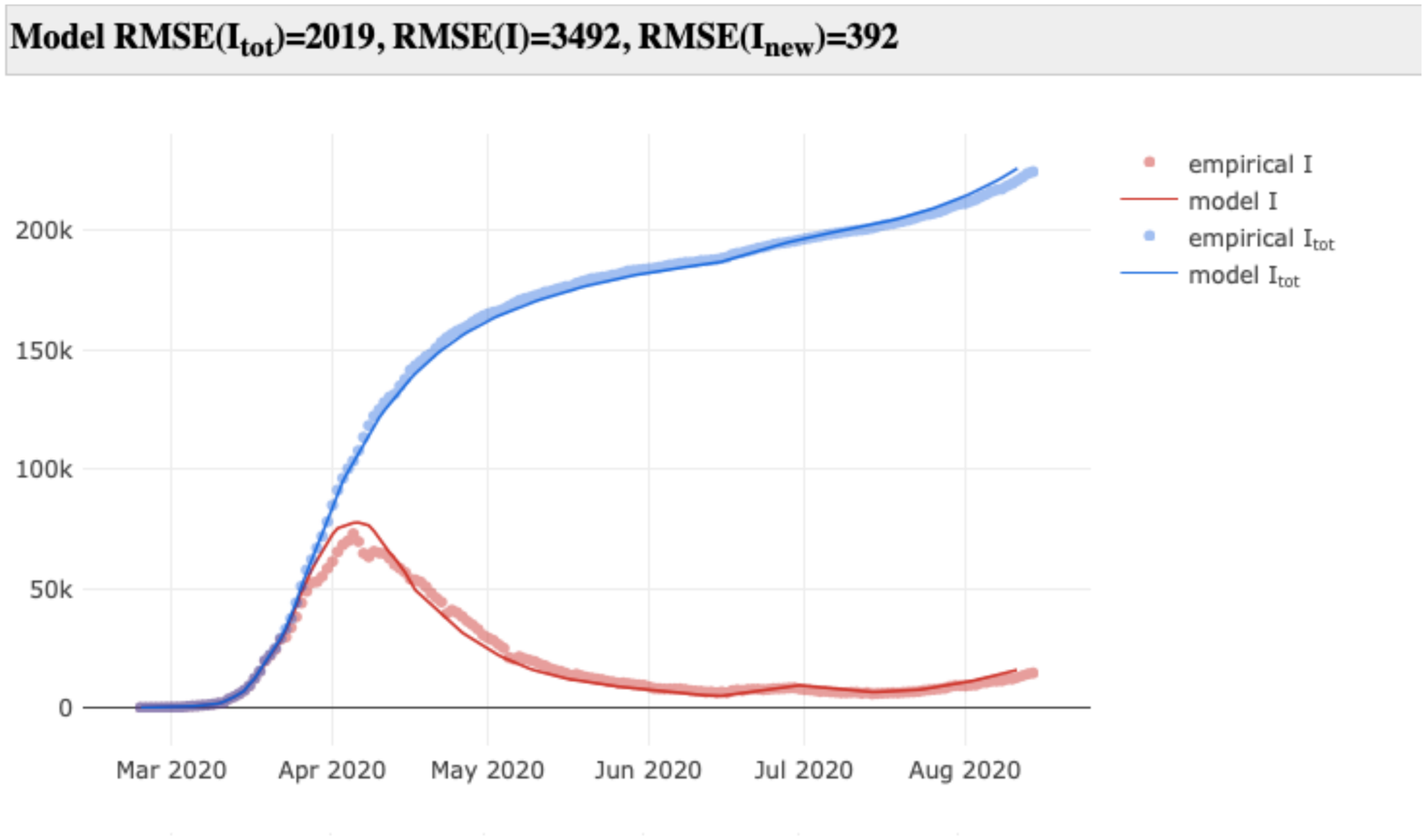} 

This leads to the question whether the SEPIR model can be useful. 
In periods where the reproduction number is nearly constant the comparison of our  model with the data curves gives a convincing picture. Assuming that this number remains constant for a while, one can predict how the development will be.   Of course, in reality the reproduction number may be considered  constant only for limited periods. 

There is a critical moment, namely when the reproduction number falls from above below 1 or passes 1 from below. This is the moment where the SEPIR model shows its full power. In Germany the reproduction number fell  below 1 at March 25 for the first time. The SEPIR model uses a constant input parameter 
for  the reproduction number between March 25  and a little bit more than a month later (see box on page \pageref{box D}). Even then  the model curve gives a  good approximation of the data as shown in the figure above (or  fig.  \ref{SEPIR D}). So, in this situation, the SEPIR model is a good approximation of the data curves and for a long period (assuming that the reproduction number is nearly constant), one obtains good predictions. We will discuss this in section 4. 


Another use which we discuss in section 4 is that the SEPIR model can be a tool for controlling an epidemic via trial and error. The ability to look into the near future via the  model allows to impose restrictions which hopefully push the reproduction number down and, after 10 - 14 days, one can check the success of these measures by applying the SEPIR model again. 

Harald Grohganz has written a program which everybody can use to compute the development of Covid-19 for a period of 4 weeks starting with an arbitrary date between April 3 and the actual day. Based on the data available 5 days before this date (based on the process of the infection about 10 days before) one can see how the  number of isolated people changes
\protect{\url{https://www.hcm.uni-bonn.de/homepages/prof-dr-matthias-kreck/modelling-epidemics/}}.

\section{The approximative SEPIR compartment model}

The idea of our model is very simple. We observe 5 compartments which we call $S$, $E$, $P$, $I$, $R$, which people pass through in this order:  {\em Susceptibles} in compartment $S$ move to compartment $E$, where they are {\em exposed} but not infectious, after they are infected by people from compartment $P$, which are {\em propagating} the virus.  From compartment $P$ they are sent to quarantine or hospital after they observe symptoms. So we call them {\em isolated} and denote this compartment by   $I$. The notation $I$ is a bit misleading, since it is often used for the infected people. What me mean is the people who are called active cases in the  worldometer form the  {\em John Hopkins University} and other data resources.  From compartment  $I$  they finally move to compartment $R$ after {\em removal } (recovery or death). Given the duration the people are members  of the different compartments, one can count the number of people in the corresponding compartments recursively and this is our SEPIR model. One might call it a delay model. The model is just a picture of what is happening in reality making the usual simplifying assumptions, in particular that averaging is allowed due to the large numbers involved. The scheme of the model is:
$$
S \to E \to P \to I \to R
$$
We call this model the {\em recursive $SEPIR$ model}. 

This schematic picture describes the model in general. 
We assume for our model that if $P(k)$ denotes the number of persons actively propagating  the virus there is a time  dependent parameter $a(k)$  called the {\em infection rate} such that we get a number of {\em additional exposed people at day $k$}, which we denote by $E_{add}(k)$:
\beq
E_{add} (k) = (1-(1-a(k-1) /N)^{P(k-1)}) S(k-1) \label{eq E-new}
\eeq 
Here $N$ is the total number of the population. This needs an explanation. For this we make the standard assumption that a single infected infects $\alpha (k)\cdot S(k)$ of the susceptible people, where $\alpha(k)$ measures the strength of the infection at day $k$.  If we have more infected, they have to ``share'' the susceptibles. For example if there are two infected the first infects $\alpha (k)\cdot S(k)$ and so the other one has only $S(k)- \alpha (k) \cdot S(k)$ susceptibles remaining. This leads to an addition of $\alpha (k)(S(k)-\alpha (k) \cdot S(k))$ infected people, which gives $(\alpha (k) + \alpha(k) - \alpha(k)^2)S(k)= (1-(1-\alpha (k))^2)S(k)$.  More generally, if there are $P(k)$ infectious people we assume inductively that $P(k)-1$ infect $(1-(1-\alpha (k))^{P(k)-1})S(k)$ people and so for the last there are only $S(k)(1- (1-(1-\alpha(k))^{P(k)-1})= S(k)(1-\alpha)^{P(k)-1}$ susceptibles remaining leading to a total of  $S(k)(1-(1-\alpha(k) )^{P(k)-1}+ \alpha(k) (1-\alpha(k))^{P(k)-1}))= S(k)(1- (1-\alpha(k))^{P(k)})$ infected people. Finally we renorm $\alpha (k)$ and replace if by $\frac {a(k) }N$, where $N$ is total number of the population. This leads to the formula above. For small $\frac {a(k)} N$ the formula is approximately equal to the simpler formula 
\beq {E_{add}(  k)}   \approx \frac{a(k-1)}{N}\, P(k-1) S(k-1) = a(k-1)\, s(k-1)\,P(k-1)\;, \label{eq E-new simplified}
\eeq
with $ s(k)=\frac{S(k)}{N}$,  which we will frequently use. At the beginning of an  epidemic the value of $s(k)$ is  $s(k)\approx 1$,   which leads to the approximative formula mentioned above:
$$
E_{add} (k) \approx a(k-1) P(k-1)
$$

We would like to stress that this is the only place in an epidemic where new people are added to compartments $E$, $P$, $I$ and $R$. Namely at day $k$ the number of exposed people is equal to the number at day $k-1$ plus the additional exposed people at day  $k$ minus those which move to the next compartment $P$. These are the people who were  exposed at day $k-e$. We express this in the formula:
\beq
E(k) = E(k-1) + E_{add}(k) - E_{add}(k-e) \label{eq E(k)}
\eeq

Similarly  at day $k$ we have additional propagating people, whose number we denote by $P_{add}(k)$, which is equal to those people leaving compartment $E$ at day $k$, which are the people who were additionally exposed at day $k-e$, so 
$$
P_{add} (k) = E_{add} (k-e).
$$ 
Like for compartment $E$ we obtain the recursive formula
$$
P(k) = 
P(k-1) + P_{add}(k) - P_{add} (k-p)=  P(k-1) + E_{add}(k-e) - E_{add}(k-e-p).
$$
The same happens with compartment $I$ leading to the formula:
$$I_{add}(k) = P_{add} (k-p)
$$ 
and using the formula for $P_{add}$ we obtain:
\beqarr
I(k) &=&  I(k-1) + I_{add}(k) - I_{add} (k-q) \nonumber  \\
 &=&  I(k-1) + P_{add}(k-p) - P_{add}(k-p-q)  \nonumber   \\
&=& 
I(k-1) + E_{add}(k-e-p) -E_{add}(k-e-p-q)\label{eq I(k)}
\eeqarr
For the compartment $R$ we apply the same principle and leave the details to the reader, but  there is a difference, since there is no other compartment to which the people move, they stay for all times. Thus the recursive formula for $R$ is:
\beq
R(k) = R(k-1) + E_{add}(k-e-p-q) \label{eq P(k)}
\eeq
Finally we recall that 
$$
S(k) = N - E(k) - P(k) - I(k) - R(k),
$$
where $N$ is the total number of the population. 

Summarizing the formulas above yields our recursive SEPIR model:\\

{\bf The SEPIR model:} {\em  Let $e$, $p$, $q$ be integers standing for the duration of staying in the corresponding compartments.    The quantities $S(k),\, E(k),\, E_{add}(k), \, P(k), \, I(k),\, R(k) $ of the SEPIR model are given by 
\begin{itemize}
\item[a)] the start condition:\\
 $E_{add}(0) = E(0) = E_0$,  a given number of exposed at day $0$,  \\ $ P(k) = I(k) = R(k) = 0$ for $k \le 0$,\\
  $E_{add} (k) = E(k) = 0$ for $k < 0$
 \item[b)]  and the recursion:
\beqa 
E_{add}(k) &=&  (1-(1-a(k-1)/N)^{P(k-1)}) S(k-1)\\
E(k) &=& E(k-1) + E_{add}(k) - E_{add}(k-e)\\
P(k) &=& 
P(k-1) + E_{add}(k-e) - E_{add}(k-e-p)\\
I(k) &=& I(k-1) + E_{add}(k-e-p) -E_{add}(k-e-p-q)\\
R(k) &=& R(k-1) + E_{add}(k-e-p-q)\\
S(k) &=& N - E(k) - P(k) - I(k) - R(k)
\eeqa

 \end{itemize}
} 

As one can see it is easy to program the model. The start condition of the recursion has to be worked out carefully.  We will say more about this later.

Before we move to a comparison of the model with data we would like to discuss the role of the parameters  $a(k)$, $e$, $p$,  $q$, which is rather different. All these parameters are a priory time dependent, but we assume for this text that $e$, $p$, and $q$ are constant. The parameter $e$ is close to a biological datum which depends on the virus, whereas the parameter $p$ results from a mixture of a biological datum and political decisions concerning quarantine regulations. 

The parameter $a(k)$ is  central for the recursion and we will derive it from the data. If one wants to apply the model for looking into the future one has to make assumptions about $a(k)$. We observe from the data, that the value for $a(k)$ fluctuates a lot during the first days. But after some time,  when the number of infected people gets large, we see periods where $a(k)$ is nearly constant  for  most countries. In these periods we actually chose $a(k)$ as a constant value. These are the periods where one can make predictions (although all predictions have to be treated with reservations). 

The choice of the parameter $q$ is a bit mysterious. To a large extent it  depends  on how good the data are reported. We observe in several countries that there are good reasons to assume that recovered or dead people are reported with a delay which one either has to reflect by choosing time dependent values for $q$ or by adjusting the number of removed people. In the countries we will discuss here in part I we don't observe this phenomenon and work with constant $q$. 

At any rate the role of $q$ is problematic and so it is good to observe that at least as long as $S(k)$ is approximately equal to the number of the population $N$  or equivalently $s(k) = S(k)/N \approx 1$ (which in all countries is fortunately the case so far) there is a function which doesn't depend on $q$.  This is the {\em total number of infected until a given day $k$ denoted by $I_{tot}(k)$}. This is the sum of $I(k)$ and $R(k)$:
$$
I_{tot}(k) = I(k) + R(k)
$$ 
Using our formulas from the SEPIR model we conclude:
$$
I_{tot} (k) = I_{tot}(k-1) + E_{add} (k-e-p), 
$$
so $I_{tot}(k)$ only depends on $a(k)$, $e$, $p$ and $s(k) = S(k)/N$, which occurs in the formula for $E_{add}(k)$. So $I_{tot}(k)$ only depends on $a(k)$, $e$, $p$ and not on $q$, if $s(k) \approx 1$. Fortunately  ${I}_{tot} (k)$ is the  datum  in an epidemic, which together with the number of  newly infected is probably best documented. More precisely, the best reported empirical data is the number of newly infected at a day $k$, which we denote by $I_{new}(k)$ and its relation to $I_{tot}(k) $ is that  $I_{tot} (k) =   I_{tot} (k-1) + I_{new}(k) $. 

To  distinguish the model function $I_{tot}(k)$ from the function occurring in the data corresponding to it, we give the the data values a different symbol and denote it by $\hat I_{tot}(k)$. In general we decorate the data values corresponding to our model functions  with a roof, e.g. $\hat I(k) $, $\hat I_{new}(k)$  and $\hat R(k)$. 

There is an obvious question for any model which  reflects the reported data only. The latter do not include information about asymptomatic infected and/or  uncounted symptomatically infected.  If such information is available it is  unreliable (first) or unknowable on principle (second).  In our model, and our data analysis, we do as if they don't exist. The following consideration is a justification for this. Suppose that there is a function $h(k)$ (like percentage of hidden infectious people) such that for any day k there are $h(k) P(k)$ people who are infectious but never show up in the data. It is not clear that the strength of infection is the same as for people with symptoms, so we take it as a separate parameter denoted by $a_h(k)$. Then the correct approximative formula for $E_{add} $ is $E_{add} (k)  \approx a(k-1) (P(k-1) + a_h(k-1)h(k-1) P(k) = (a(k-1) + a_h(k-1)h(k-1)) P(k-1)$. Thus the hidden people just lead to an enlargement of the strength of infection and the model works with parameter  $a(k-1) + a_h(k-1)h(k-1)$. Thus we can do as if these people don't exist. 

But there is another parameter, namely $p$, which might be influenced by the asymptomatically infected people and their strength of infection. One might wonder, whether the SEPIR model can be used to shed some light on this question. 
 As mentioned above $\hat I_{new}(k)$ seem to be the best documented empirical data on an epidemic, so it plays a central role. We derive the most fundamental input into our model, $a(k)$, from the $\hat I_{new}(k)$. We explain this in section 3.

An important parameter for all models is what is called the {\em reproduction number}. This is the number  of people that one infected person will pass the virus on to, on average. Since we assume that only people in compartment $P$ are propagating the virus and each day a single person in compartment $P$  infects approximately $a(k)$ people, we identify the reproduction number in our model as $\mathcal  {R}(k) := p\, a(k)$.

\section{Reconstruction of the parameters from the data}

Although we are convinced that already the assumptions on which the SEPIR model are based are closely related  to reality and so it has a good chance to model the reality, the final test is, as with all models, the comparison with the data. The challenges which are posed to all models are the following:
\begin{itemize}  
\item [--] Find a method how to derive the (time dependent) input parameters from the data set.
\item [--] Find time intervals where the input parameters are nearly constant  and  check whether the model curves are approximatively equal to the data curves. The longer these intervals are the more useful the model is, for these are the periods where one can use the model for predictions. 
\end{itemize}

We assume $e=4$ and $p=5$ as  constant durations for all countries we deal here with, although one may also like to consider different values  and to check the consequences for the model predictions. We plan do to so in a forthcoming paper \citep{GKS}. The number $e$ is a medical datum, $p$ is the sum of the mean duration of infectiousness without symptoms, estimated as 2 days, and the average time delay between the observation of symptoms and the beginning of the quarantine, which for the chosen countries should not be longer than about 3 days.\footnote{Compare \citep{QunLi_ea:2020} and
 \protect{\url{https://www.rki.de/DE/Content/InfAZ/N/Neuartiges_Coronavirus/Steckbrief}} .}
 This leads us to the estimated value 5 for $p$. A check of the model performing approximations for Switzerland and South-Korea with different values for $p$ indicates  an optimum of  the approximation for  $p$ between 4 and 6. In addition  we tried out smaller values for $e$ and found that this makes the model unstable in respect to the optimization (described below) of the intervals, where $a(k)$ is approximatively constant.  And making $p$ large leads to unnatural fluctuations. In the light of the general quarantine regulations in many countries  and  the overall (numerical) dominance of weakly infected persons, we expect   $q$ to be about $14$ or $15$ but check the reliability of this expectation in an initial analysis of the  data. In some countries it turns out to be larger and in 
  some cases even time  dependent. We will discuss the choices of the parameters $e$, $p$, and $q$ more carefully in future work.

 The most delicate and central input of our model are the parameters $a(k)$.  They are derived from the daily reported number of newly infected $\hat I_{new}(k)$. We actually replace the data $\hat I_{new}(k)$ by the 3 days average, which we call $\tilde I_{new}(k)$. The effect is that we reduce the sometimes enormous jumps of the number of newly infected. The $\tilde I_{new}(k)$ correspond to  the $I_{new}(k)$ in our model. The model tells how to derive the parameters $a(k)$ from the numbers $\tilde I_{new}(k)$. As before we give the numbers which result from the data a different name and call them $\bar a(k)$. 
 
 Because of the time delay between the infection at  day $k$, resulting in $E_{add}(k)$ and its visibility in the number of registered infected,  the infection is expressed by  the   number of registered newly infected $I_{new}(k+e+p)$. The number of actively infectious persons at $t=k$,  i.e. those in the compartment $P$ at this day, is the sum of persons who have themselves been newly infected at a  day $k-(e+j)$ for $j= 0, \ldots, p-1$. They  appear   $p+e$ days later as visibly new infected, i.e. as $I_{new}(k+p-j)$.  Using $\tilde I_{new}(k)$ instead of $I_{new}(k)$ in our model the equations of the model determine the values for $\bar a(k)$ (under the assumption $s(k)= \frac{S(k)}{N} \approx 1$): 
\beq \bar{a}(k) := \frac{\tilde{I}_{new}(k+e+p)}{\sum_{j=0}^{p-1}\hat{I}_{new}(k+p-j)} \; . \label{eq a(k)}
\eeq 

Now we begin with the determination of the time intervals. We fix the first day from which on we determine $\bar a(k)$ and call it $t_0 = 1$ as the day where for the first time the number of newly infected people $\hat I_{new}(k)$ is $>0$ for at least $e+p$ days. 
As an effect of the restrictions imposed by governments, $\bar a(k)$  starts to decay  after a few days   and, hopefully,  falls until it reaches  a  value approximately equal to $0.2$. Assuming $p=5$ this corresponds to a reproduction number $\mathfrak R(k) = 5 \,\bar a(k) \approx 1$. We denote the first day where $\bar a(k) \le 0.2$ by $t_1$.   For many countries one can observe  that after day $t_1$, where the reproduction number is approximatively equal to $1$, periods of considerable length follow, for which $\bar a(k)$ is approximately constant. These periods  are modelled by intervals $A_1 =[t_1,t_2)$, $A_2 = [t_2,t_3)$, $A_3 = [t_3,t_4)$ and so on, which we consider as our {\em main intervals}. In in these  periods  our model can work with constant values of the $a(k)$ determined from the $\tilde {I}_{new}(k)$ in a systematic way. 

The determination of the main intervals is a delicate step which we will comment upon later. Before we do so, we explain how we treat the interval $[t_0,t_1)$. Since our model is a recursive one we need  values for $E_{add}$  for at least $e+p$ days before the day $t_0$  to get the model started. These $e+p$ days play the role of an electric engine starter, which is needed at the beginning until the engine runs smoothly. We choose the first $e+p$ values of $\hat{I}(k)$ starting from $k=1$ and shift them $e+p$ days back.   If we set all the $a(k)=0$ for $k<0$ and consider $a(0)=a_0$ as a free parameter this gives the necessary information to start the model recursion at $t_0=1$ for any value of $a_0$. 
 In the end we look for the value of $a_0$ for which the resulting model curve for $I_{tot}(k)$ using the daily values for $\bar a(k)$  comes closest to the empirical curve $\hat I_{tot}(k)$; closest in the sense of minimizing the distance between $I_{tot}$ and $\hat{I}_{tot}$, using the {\em root mean square error} $\textsl{RMSE} (I_{tot})$ as the criterion.
   Since we have here a single free parameter this can easily  be found either by computer or by hand. This determines our {\em start value}. We have chosen   $\textsl{RMSE}\, (I_{tot})$ for the measure since this is independent of the more  problematic parameter  $q$. Once we know the start value we model the interval from $t_0$ to $t_1$ using the daily values for $\bar a(k)$. Using the daily strongly varying values for $\bar a(k)$ in this interval and no longer intervals implies that in the interval between $t_0$ and $t_1$ model predictions are impossible. 
   

Now we discuss the determination of the main intervals starting from $t_1$. The hope is to find a small number of main intervals which still leads to a good approximation. The basic idea is the following. Fix a number $m$ of main intervals and minimize $\textsl{RMSE}\,(I_{tot})$ under this condition. There are standard methods how to solve such an optimization problem.  But there are simple by hand methods, where the number $m$ is left open. First look at the curve for $\bar a(k)$ and check whether there are points, where the curves makes obvious bigger jumps. Chose a point close to each of these obvious jumps and call them ``jump points''. Compute the model curves and the $\textsl{RMSE}\, (I_{tot})$ using the intervals given by the jump points, the beginnig $t_1$ and the end of the period you would like to model. Now divide one after the other the resulting intervals in the middle, unless their length is less than 10. For each new interval end compare the $\textsl{RMSE}\, (I_{tot})$ with the previous value. If this is almost the same, remove the new interval end. Otherwise move the middle point to the left or right until the $\textsl{RMSE}\, (I_{tot})$ reaches the minimum. Now consider the next interval and do the same and so on. At the end make a fine tuning by moving the interval points a bit until one reaches a minimum for $\textsl{RMSE}\, (I_{tot})$.

\section{Modelling  Covid-19 for Germany and Switzerland}
We demonstrate this ad hoc method by looking at Germany and Switzerland. The data we are using stem from John Hopkins University (JHU).\footnote{For most countries we use the humdata repository of JHU directly  \url{https://data.humdata.org/dataset/novel-coronavirus-2019-ncov-cases}. For  Switzerland we use the corresponding time series contained in the  {\sc Mathematica}  resources for  Covid-19. They agree basically with the JHU data; for information see 
  \protect{ \url{https://datarepository.wolframcloud.com/resources/Epidemic-Data-for-Novel-Coronavirus-COVID-19}}. } 
  
\begin{small}
\vspace{1em}
\begin{center}
\begin{tabular}{|c|c|c|}
\hline 
\multicolumn{3}{|c|}{\textbf{Basic parameters D, CH}}\\
\hline
$e$ & $p$ & $q$   \\
\hline 
4 & 5 & 15
\\
\hline
\end{tabular}
\end{center}
\end{small}
  
\pagebreak

\begin{figure}[h]
\includegraphics[scale=0.4]{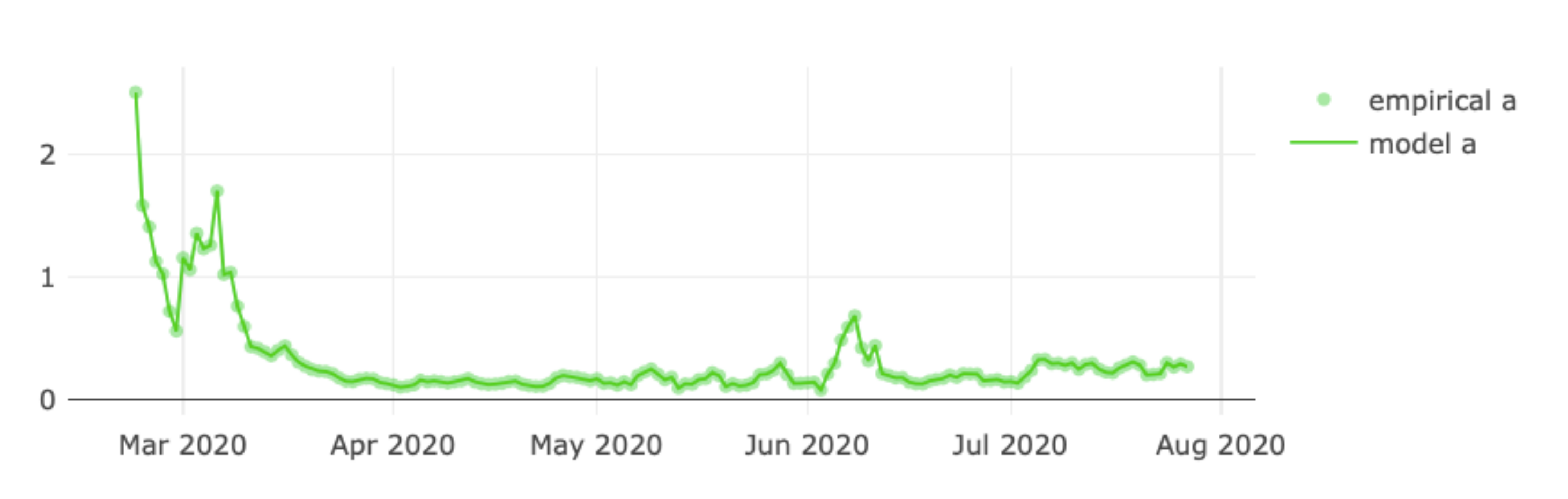}
\vspace{-0.1cm} 
\caption{\tiny Infection rates $\bar{a}(k)$ for Germany,  Mar 25, 2020, is the first day $t_1$ with infection rate below $0.2$. It is parametrized by $k=30$;  here $\bar{a}(30)=0.158$. \label{fig a(k) D}}
\end{figure}

For {\em Germany} (D) the day $t_0$ at which the number of newly infected becomes non-sporadic, i.e. without interruptions of days with no newly infected, is  Feb. 25, 2020.
 Here we start our count of days, $t_0=1$. The $\bar{a}(k)$ values for Germany, determined according to equation (\ref{eq a(k)}), are shown in figure \ref{fig a(k) D}. 

\begin{figure}[h]
\includegraphics[scale=0.4]{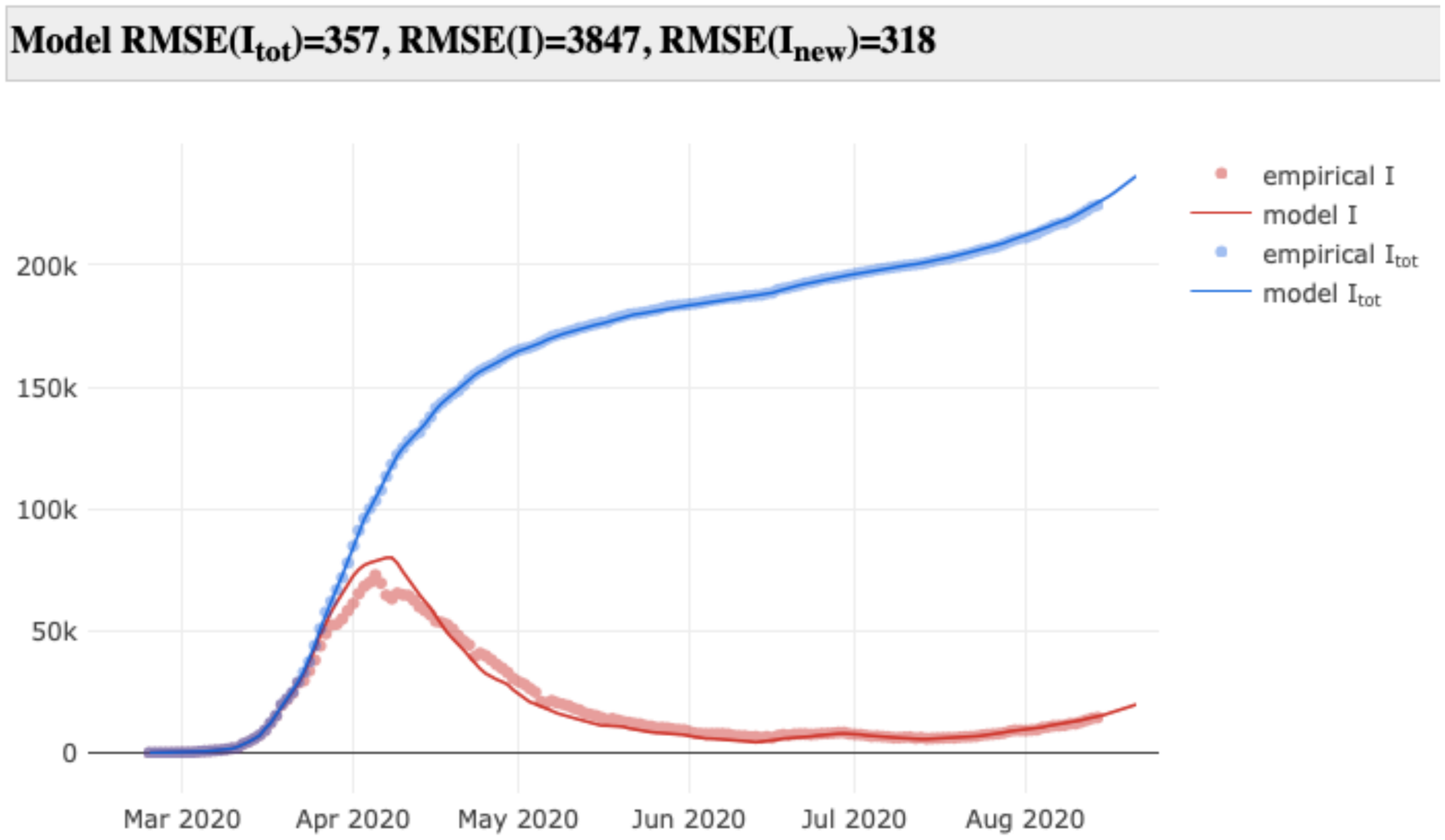}
\caption{\tiny SEPIR model curves (unbroken lines) and JHU data (dots)  for the confirmed cases $I_{tot}$ and the actually infected $I$ for Germany. Model curves here determined with daily changing $a(k)=\bar{a}(k)$ for optimal start condition explained in the main text with $a_0=2.665$. In the headline the root of mean square difference (RMSE) of the $I_{tot}$, $I$ and $I_{new}$ (not depicted) to the corrsponding JHU values are given. \label{fig Itot tagesgenau D}}

\end{figure}

The first day with $\bar{a}(k)\leq 0.2$ is March 25, $t_1=30$; the end of data for the following  consideration is  $k=170$ (August 12, 2020). The optimal start value for Germany is $a_0 = 2.665$. With the  start condition explained above the recursion with day by day changing model values according to the empirical data, $a(k)=\bar{a}(k)$, leads to a result shown in figure \ref{fig Itot tagesgenau D}.

\pagebreak
If we let the  total period end at day $k= 170$, one sees just two obvious jump points at $k=132, \, 142$, i.e. around June 8, this is about 9 days before  a hotspot in a German meat factory  became apparent in the $\hat{I}_{new}$ , so that about 7000 employees and their families had to be sent  into quarantine for two weeks (the T\"onnies case).  
With these four values $k=30, \, 132, \, 142,\, 170$ the $\textsl{RMSE}\, (I_{tot})$ becomes $19548$. Now add the new interval end 66; then $\textsl{RMSE}\, (I_{tot}) = 9209$. Moving this end to the right gives a worse value, while moving it to the left gives improvements each time until the optimum is reached at 62, where  $\textsl{RMSE}\, (I_{tot}) = 7516$. Now introduce a new interval  end $141$, then $\textsl{RMSE}\, (I_{tot}) =5387$. Moving this point to the right gives worse values, while moving it to the left improves the values each step until an optimum is reached for 132 with $\textsl{RMSE}\, (I_{tot}) =5071$. A fine tuning by slightly moving the interval ends  leads to the interval ends $30,\, 65,\, 103, \,110, \,132, \,170$ with $\textsl{RMSE}\, (I_{tot}) = 2019$. 

The   dates of the partition points of the time-line are given in the following table.  The mean values $a_j$ of the $\bar{a}(k)$ in the respective intervals $[t_j, \, t_{j+1})$ and the empirical standard deviation follow.
\vspace{2em}

\begin{small}
\begin{center}
\vspace{0.5em}
\begin{tabular}{|c|c|c|c|c|c|c|}
\hline 
\multicolumn{7}{|c|}{\textbf{Time partion SEPIR model Germany}}\\
\hline
$t_0\; (k=1)$ & $t_1\; (k = 30)$ & $t_2\; (k=65)$ & $t_3 \; (k=103)$ & $t_4 \; (k=110)$ & $t_5\; (k=132)$ & $t_6\; (k=170)$ \\
\hline 
02/25 & 03/25  & 04/29 & 06/06 & 06/13 & 07/05 & 08/12 
\\
\hline
\end{tabular}
\end{center}

\vspace{0.5cm}
\hspace*{-0.5cm}\begin{tabular}{|c|c|c|c|c|}
\hline 
\multicolumn{5}{|c|}{\textbf{SEPIR infection rates $a_i\, (\pm \sigma)\; (\mathtt{ standard\, deviation}$) in main intervals for Germany}}\\
\hline
  $a_1$ & $a_2$ & $a_3$ & $a_4$ & $a_5$\\
\hline 
$0.139\, (\pm 0.024)$ & $0.160\, (\pm 0.047)$& $0.459\, \pm (0.140)$ & $ 0.167\, (\pm 0.028)$) & $0.260 \, (\pm 0.038)$\\
\hline
\end{tabular}
\label{box D}
\end{small}

\begin{figure}[h]
\includegraphics[scale=0.4]{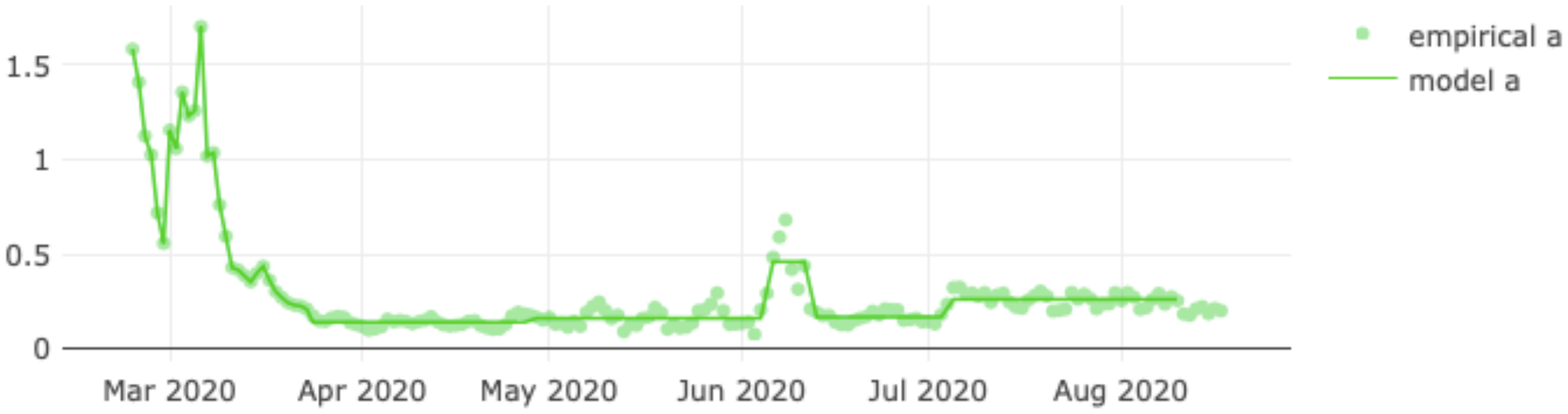}
\vspace{-0.1cm} 
\caption{\tiny Infection rates $\bar{a}(k)$ for Germany and model values $a_j$ in the main intervals (see table)\label{fig a(k) D mit Modellwerten}}
\end{figure}

\newpage
 Figure \ref{fig a(k) D mit Modellwerten} shows the model values $a_j$ in the main intervalus in comparison with the the daily values of the $\bar{a}(k)$.
 
\begin{figure}[h]
\includegraphics[scale=0.4]{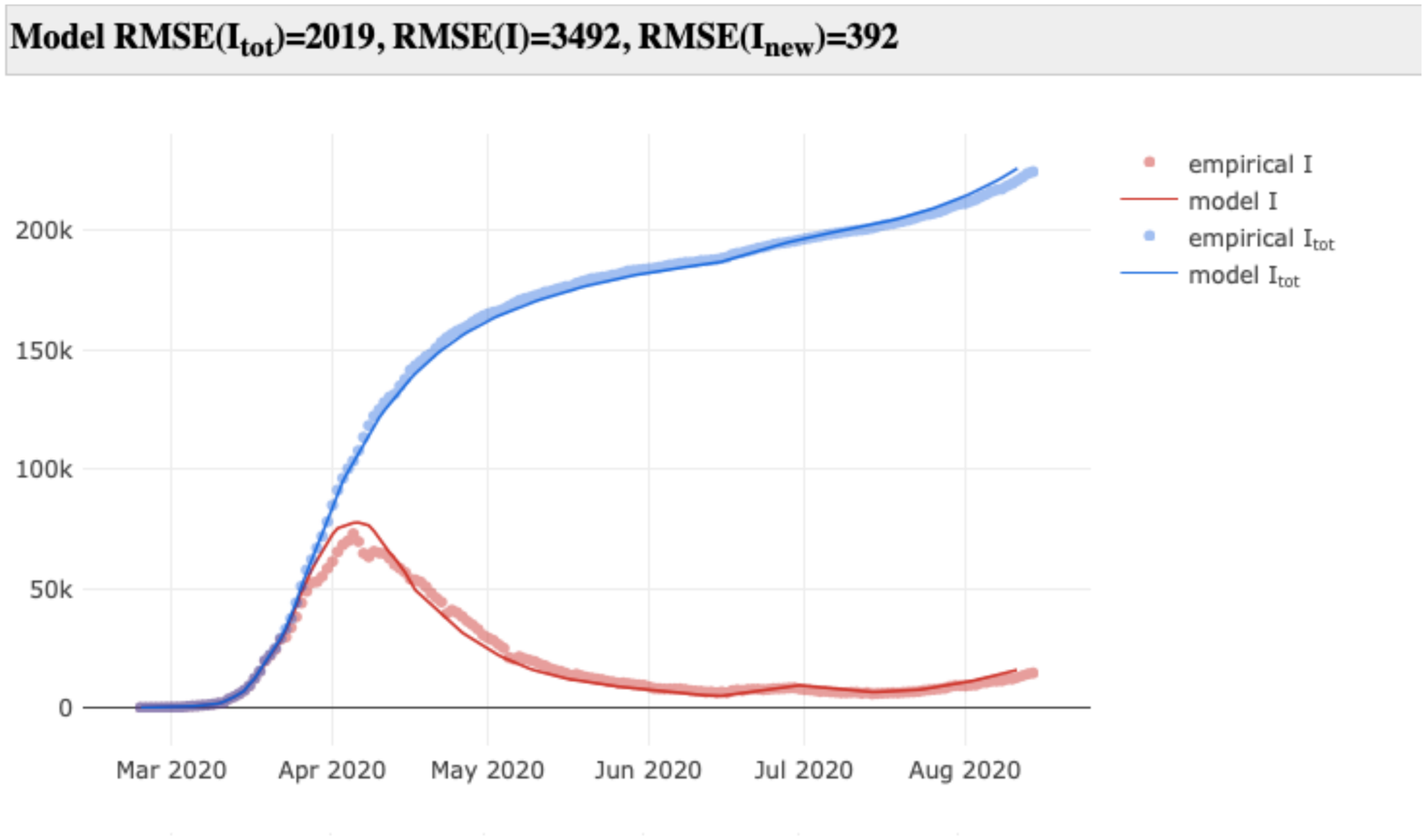}
\caption{\tiny The SEPIR model for Germany with $a_0=2.665$ and main interval ends $30,\,65,\, 103, \, 110, \, 132,\, 170 $  starting with  $t_1=30$ (Mar 25, 2020). $I_{tot}$ blue, $I$ red;  continuous lines model values, (fat) dots JHU data.  \label{SEPIR D}}
\end{figure}

 Figure 
\ref{SEPIR D} shows a picture of the resulting model curves for $I_{tot}$ and $I$. 

\begin{figure}[h]
\includegraphics[scale=0.4]{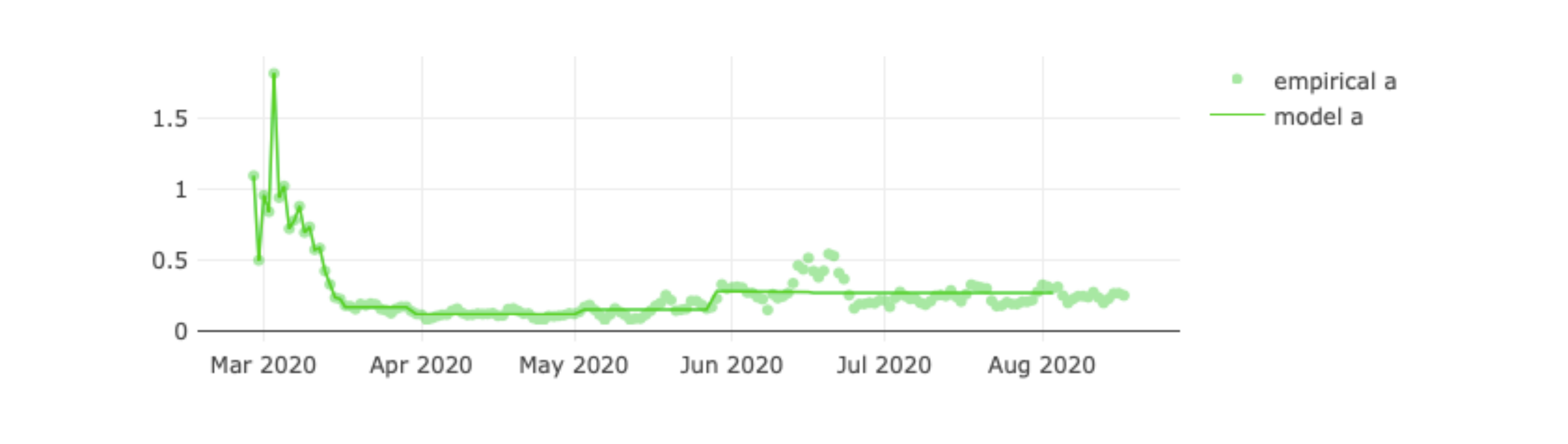}
\caption{\tiny   Infection rates $\bar{a}(k)$ for Switzerland,  Mar 29, 2020, is the first day $t_1=28$ with infection rate below $0.2$ with   here $\bar{a}(28)=: 0.491$.   \label{fig a(k) CH}}
\end{figure}

If one applies the same method to {\em Switzerland} (CH) one finds
that  the conditions for  $t_0$  are satisfied February 29, 2020, while  March 18 is the day $t_1$, i.e. $t_1=19$. The values for the $\bar a(k)$ (including the model values $a_j$ in the main intervals) are shown in figure \ref{fig a(k) CH}.

\begin{figure}[h]
\includegraphics[scale=0.3]{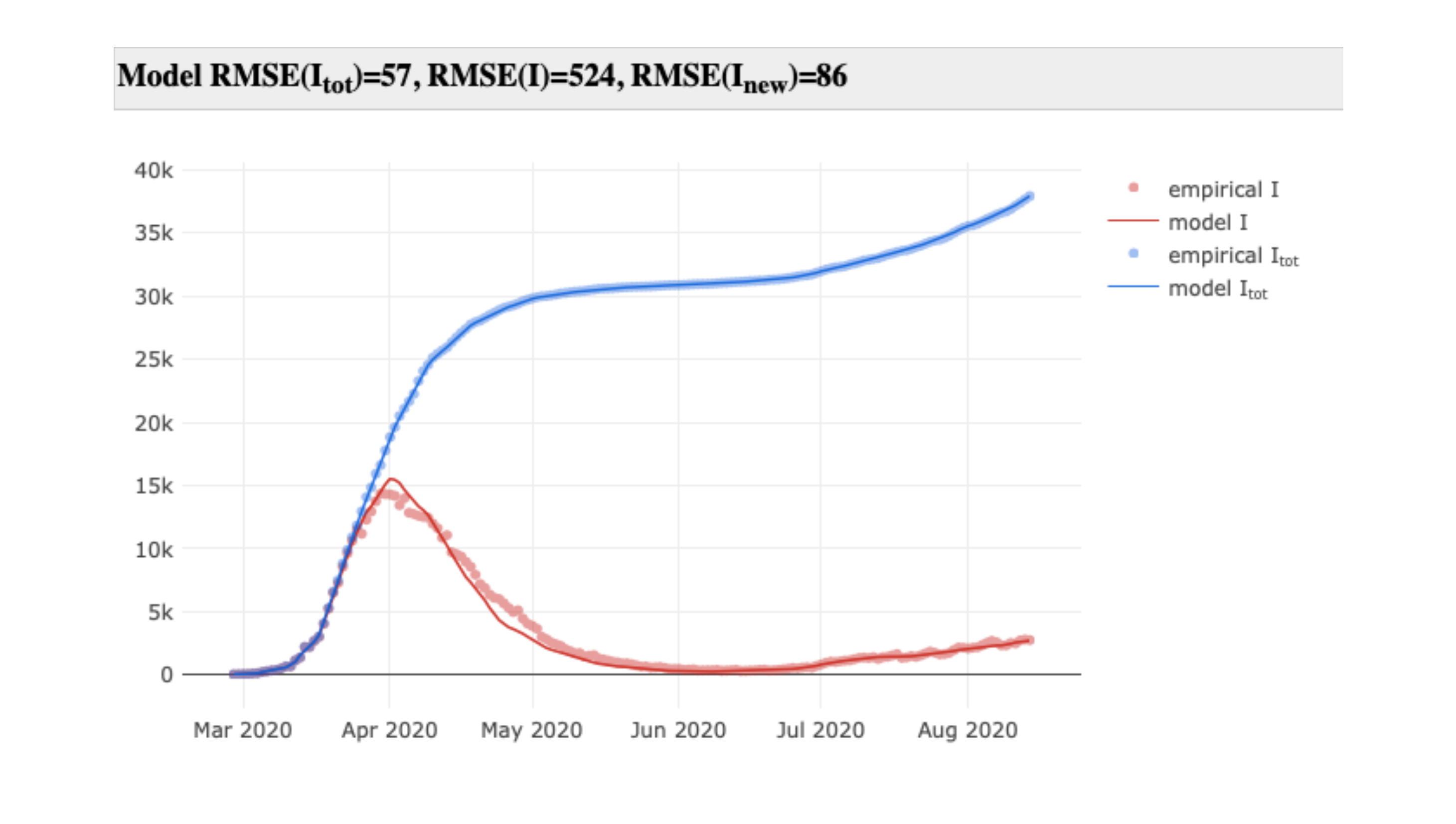}
\caption{\tiny  SEPIR model curves (unbroken lines) and JHU data (dots)  for the confirmed cases $I_{tot}$ and the actually infected $I$ for Switzerland. Model curves are here determined with daily changing $a(k)=\bar{a}(k)$ for optimal start condition explained in the main text with $a_0=0.491$.   \label{fig Itot tagesgenau CH}}
\end{figure}

 The optimal 
 start value for the recursion is $a_0= 0.491$.   
The  resulting model curve with daily values $\bar a(k)$  are shown  in figure \ref{fig Itot tagesgenau CH}.

Here it is not so obvious to find jump points, but one might make a first try with $k=$ 19, 35, 71, 99, 131, 159. Playing for a while yields a good approximation with jump points at $k=$  19, 32, 65, 91, 110, 159.
The resulting time partition and the model values $a_j$ in the  intervals $[t_j, \, t_{j+1})$ are shown in the following tables.

\begin{small}
\vspace{0.5em}
\begin{center}

\hspace*{-0.2cm}\begin{tabular}{|c|c|c|c|c|c|c|}
\hline 
\multicolumn{7}{|c|}{\textbf{Time partition  SEPIR Switzerland}}\\
\hline
$t_0\; (k = 1)$ & $t_1\; (k = 19)$ & $t_2\; (k=32)$ & $t_3 \; (k=65)$ & $t_4 \; (k=91)$ & $t_5\; (k=110)$ & $t_6\; (k=159)$ \\
\hline 
  02/29 & 03/18& 03/31  & 05/03 & 05/30 & 06/17 & 07/27 
\\
\hline
\end{tabular}

\end{center}

\vspace{0.5cm}

\hspace*{-0.9cm}\begin{tabular}{|c|c|c|c|c|}
\hline 
\multicolumn{5}{|c|}{\textbf{SEPIR infection rates $a_i\, (\pm \sigma)\; (\mathtt{ standard\, deviation}$)  in main intervals for Switzerland}}\\
\hline
  $a_1$ & $a_2$ & $a_3$ & $a_4$ & $a_5$\\
\hline 
$0.164\, (\pm 0.020)$ & $0.113\, (\pm  0.20)$& $0.147\, \pm (0.046)$ & $ 0.277\, (\pm  0.077)$) & $0.265 \, (\pm 0.90)$\\
\hline
\end{tabular}

\end{small}

\begin{figure}[h]
\includegraphics[scale=0.3]{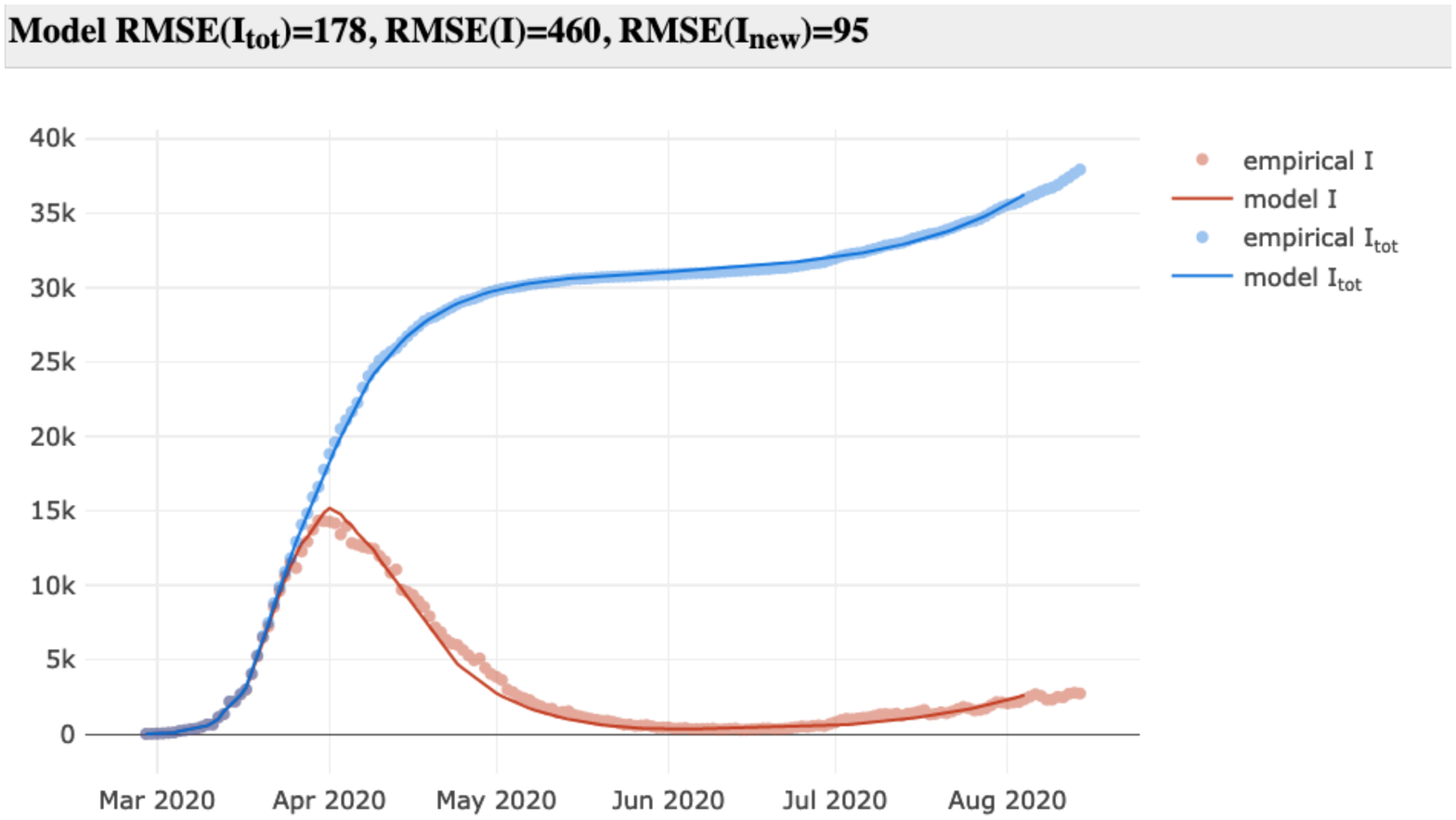}
\caption{\tiny The SEPIR model for Switzerland with $a_0=0.491$ and main interval ends  28, 44, 80, 108, 140, 168  starting with  $t_1=28$ (Mar 29, 2020). $I_{tot}$ blue, $I$ red;  continuous lines model values, (fat) dots JHU data. \label{SEPIR CH}}
\end{figure}

\newpage
 The resulting model curves  are shown in 
  (figure \ref{SEPIR CH}).


It is likely that one can obtain better results by applying optimization algorithms.  But the results of this ad hoc method are good enough to demonstrate the use of the SEPIR model.

Summarizing we have clear methods how to read off intervals where the infection rate $\bar a(k)$ measuring the strength of the infection is approximately constant and to derive the value of $a(k)$ during these intervals from the data. This can be used to model the epidemic of the different countries.

\section{Use of the $SEPIR$ model}

\subsection{Model based views into the future}

The first question people want to know from a new model is
what it tells about the future.
The natural answer is: That depends on how the people behave. The next question then may be: 
Assume that they have behaved more or less the same in the near past;\footnote{As one often hears from epidemiologists the data of a given day reflect the infection process abut 10 -- 14 days earlier. In our notation this is the period  $e+p$ days before the actual date, since  only after this many days exposed persons develop  symptoms and are counted as infected.} 
 what does the model  predict, if people don't change their  behavior in  the (near)  future?
 

 For example if we consider the data curve  shown  for Germany  above and ask this question around March 15, then the answer 
 is clear: The assumption that the people have behaved the same in the near past would not be realistic. For about 9 days before March 15 the values for $\bar a(k)$ change dramatically day by day (fig. \ref{fig a(k) D}). 
 
\begin{figure}[h]
\center 
\includegraphics*[scale=0.5]{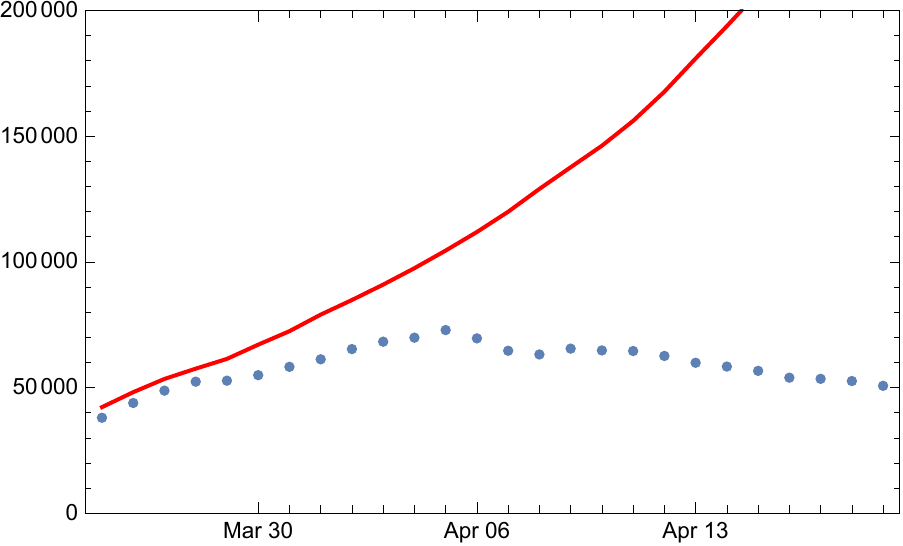}
  \includegraphics*[scale=0.5]{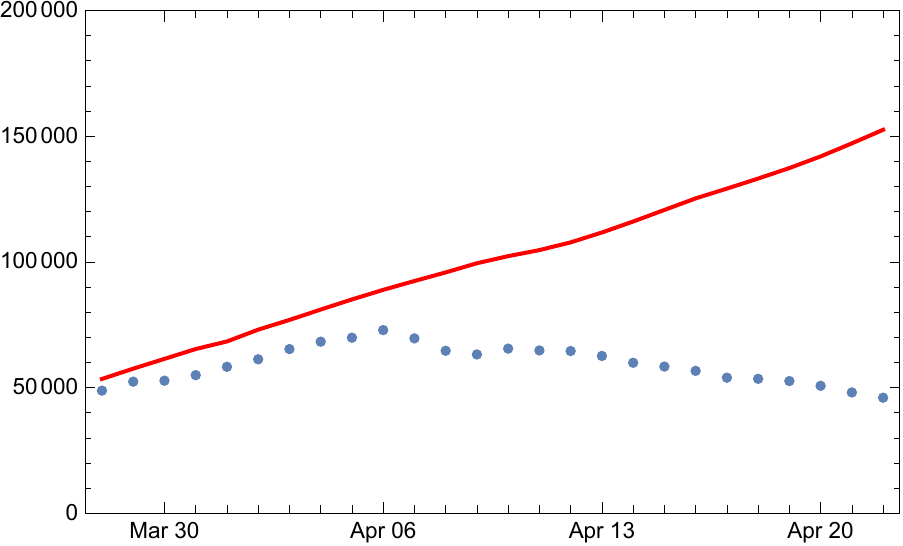}
   \includegraphics*[scale=0.5]{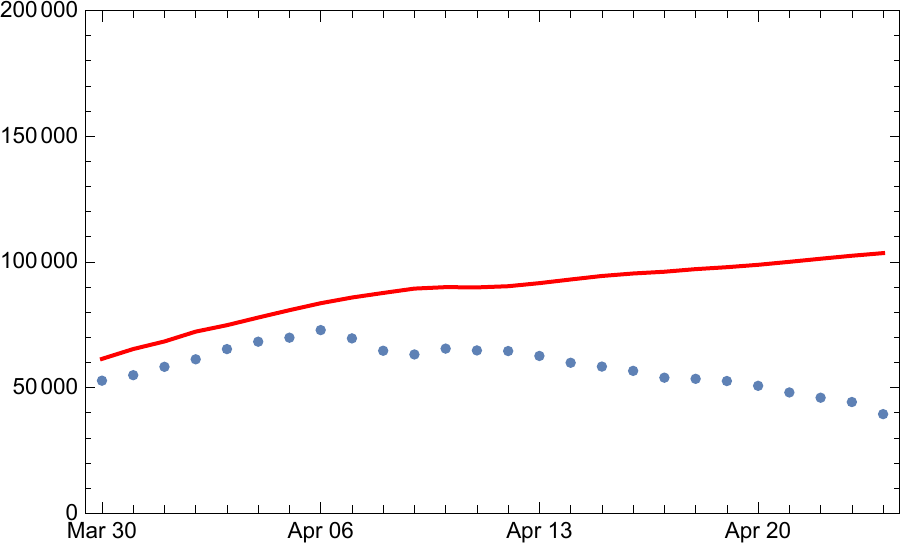} \\ 
   \center 
  \includegraphics*[scale=0.5]{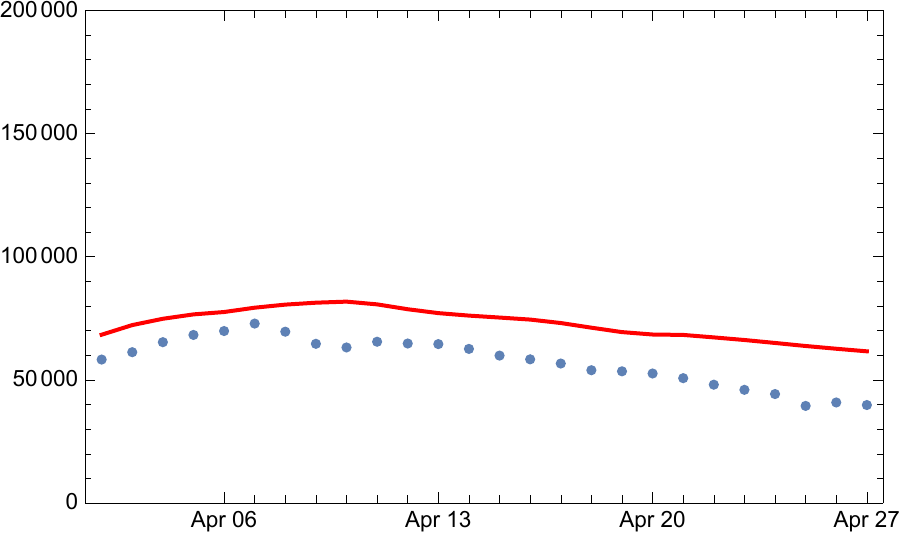}
   \includegraphics*[scale=0.5]{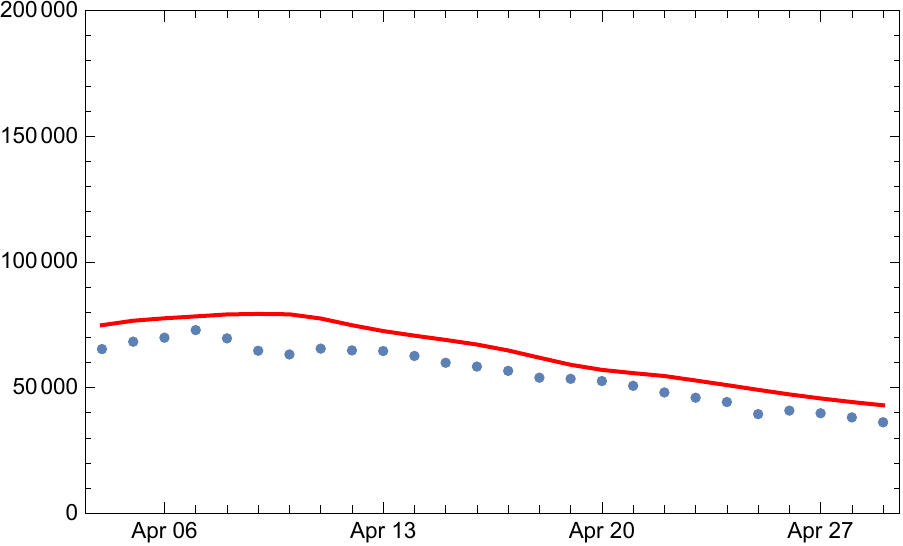}
      \includegraphics*[scale=0.5]{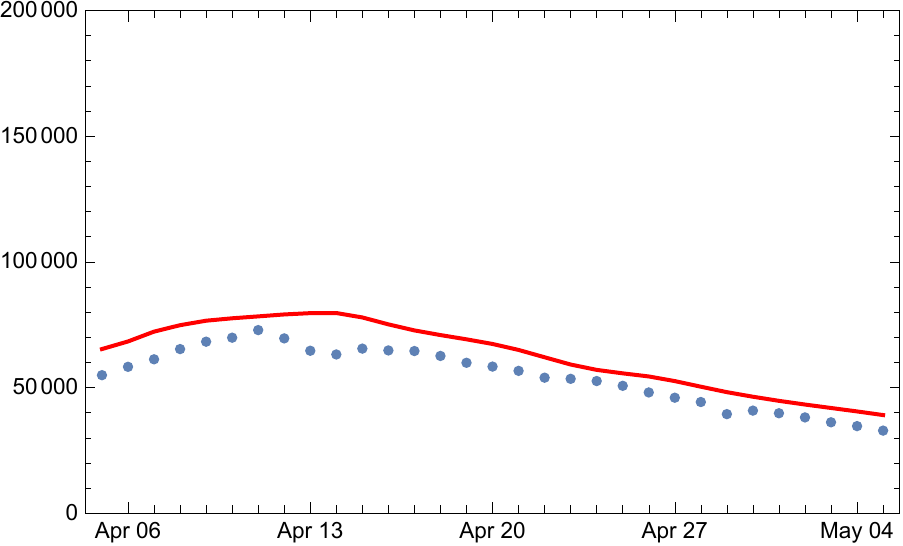}
   \caption{\tiny  SEPIR pedictions for $I(t)$  on the basis of the data for $\hat{I}_{new}$ of 5 days ending in dates $t_a$;  top: $t_a =$ Mar 31,  Apr 02, Apr 04;  bottom 
   $t_a =$ Apr 06,  Apr 08, Apr 10; blue dots JHU data $\hat{I}(t)$ \label{kinemat series}}      
     \end{figure}

 If the people would have asked this question in the first days of April, on the other hand, things would have been different. The situation accessible to  the observational data at this date (i.e., the values of  $\bar{a}(k)$ for the period  $e+p=9$ days earlier)  started to be sufficiently stable for daring a  model based prediction.
They would contain the data on  March 25, 2020  the day $t_1=30$ at which the reproduction number for Germany fell below 1 for the first time ($\bar{a}(t_1)=0.173$).  9 days later, i.e. on April 3, this became visible in the $\hat{I}_{new}(k)$; and it stayed so for the following days (in fact until May 25). If one would have  dared a provisional look into the future  with SEPIR based on the data $\bar{a}(k)$ for the three days starting with $t_1$ (April 3 -- 5 with $k=39$ to $41$), all of them below $0.2$, the SEPIR model curve    would clearly indicate  that the $\hat I(k)$-curve was  already close to the peak value, which one could expect  in the next few days (figure \ref{kinemat series}). 

So the model prediction is not so far away from the data curves. As said above the data $\bar a(k)$ values for March 24-26 are only reflected in the data about 10 day later, April 3 - 5. This is almost the time, when the $\hat{I}$ curve reaches the peak. But even at that day it was not obvious from the development of the $\hat I$-curve that the peak would be reached so soon.

\subsection{The $SEPIR$-model as a tool for trial and error} 

\begin{figure}[h]
\includegraphics[scale=0.3]{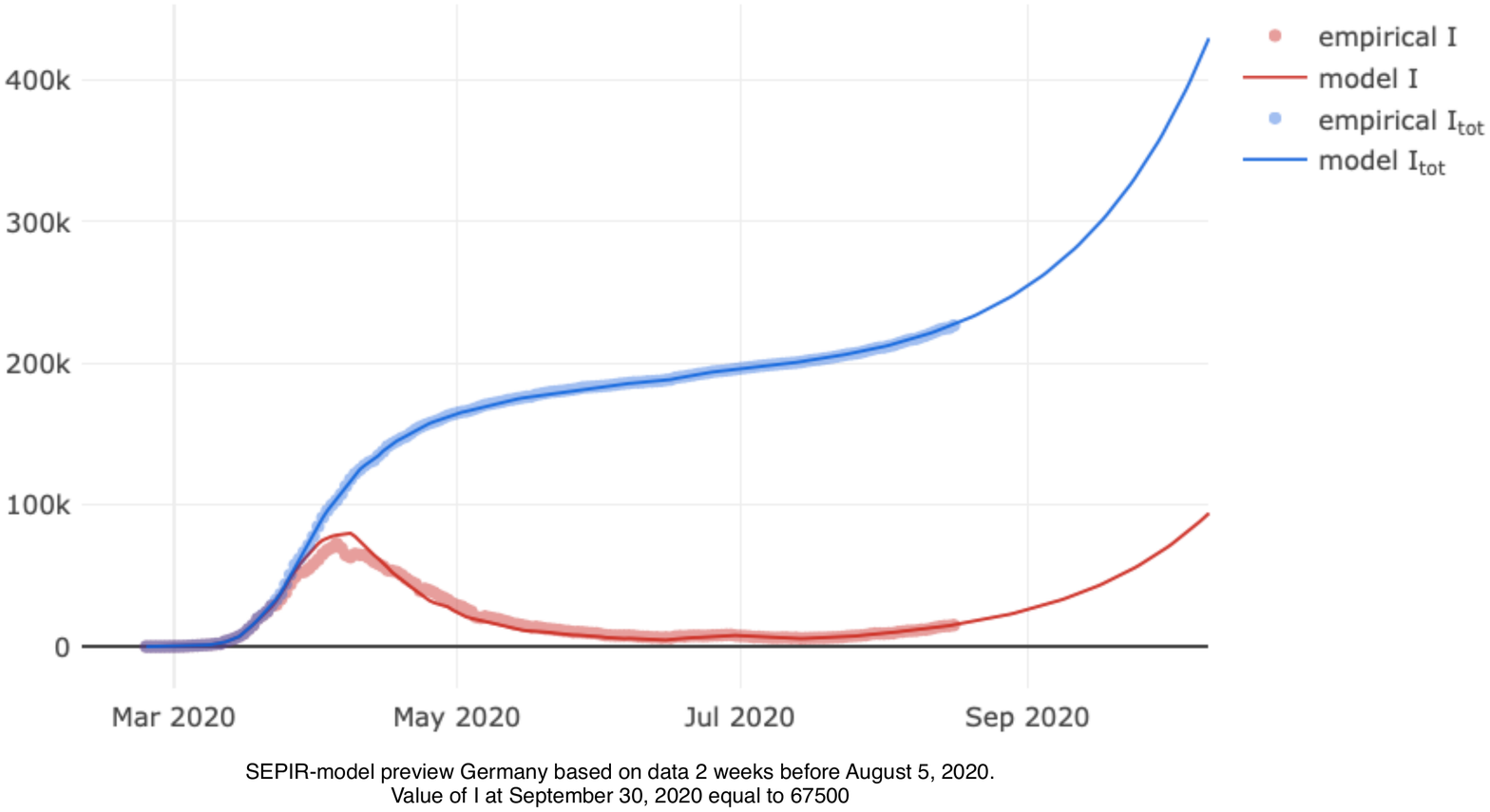}
\caption{SEPIR preview  computed on the basis of data available 2 weeks before August 5, 2020   \label{SEPIRpreviewAug5}}
\end{figure}

The natural way to control a process is to use all information available to estimate the effects of a restriction in advance. Even if one can learn from the past 
to get some feeling for  the effects of restriction  there will remain a great uncertainty. In such a situation one will apply trial and error -- if there is enough time. This was not the case in March 2020, since, besides having no experience which measure has which effect, the development was too fast. But in the future there is, generally speaking, more control  because the numbers increase slower. For example in Germany the reproduction number is above 1 since July 5, leading to a second wave if not stopped by exterior means. But the increase is much slower (SEPIR reproduction rate $\approx 1.3$ in comparison to $R\geq 2$ in March).
\newpage
Figure  \ref{SEPIRpreviewAug5} takes a look into the future based on the data during 2 weeks before August 5, 2020.

According to this prediction    the number of actually infected would  be almost 68000 by the end of September 2020, if no additional restriction measures are imposed in late August, early September. This number is close to the peak at the beginning of April, which was reached within about 3 weeks.  The data on mid August, available at the end of the month, seem to  indicate  a more relaxed situation (consult the website given below 
and check the dates August 28 -- 31).

So trial and error is possible, even if the effects of a restriction  are only visible in the data about 10 days later. Combined with some knowledge about the influence of restrictions this might allow politicians to look at the future using the SEPIR model based on the present reproduction number, impose a restriction which based on experience in the past lowers the reproduction number to a level desired. After 10 days one will see the effect and using the  model again one can take a view into the future. If like at the beginning of April this shows a quick reach of the peak, the restriction was successful, otherwise there is enough time to impose more restrictions. 

Harald Grohganz has written a program which everybody can use to compute the development of Covid-19 for a period of 4 weeks starting with an arbitrary date between April 3 and the actual day. Based on the data available 5 days before this date (based on the process of the infection about 10 days before) one can see how the  number of isolated people changes; see 
\protect{\url{https://www.hcm.uni-bonn.de/homepages/prof-dr-matthias-kreck/modelling-epidemics/}}.
\\

\pagebreak
In the final phase of our work we came across the paper \citep{Balabdaoui/Mohr:2020} which, independently, follows a similar approach. The authors also use a recursive model  with additional compartments and a stratification into different age layers adapted to the Swiss context. Our model is simpler and can be adapted to different contexts more easily.  For the Swiss case the basic results of the models seem to be sufficiently close to  trust our simple approach. 

We would like to thank Harald Grohganz for invaluable help programming our different approaches quickly, and also for several helpful comments. We are very grateful to Odo Diekmann. He was willing to discuss our thoughts as non-experts and helped us to understand compartment models better, gave hints  and corrected mistakes here and there. Finally we would like to thank Stephan Luckhaus and Viola Priesemann for stimulating exchanges.

 \vspace{5em}
\addcontentsline{toc}{section}{\protect\numberline{}Bibliography}
\small
 \bibliographystyle{apsr}


\end{document}